# An intriguing interpretation of Cosine beams

A. Srinivasa Rao[1,2,3*]

[1]Graduate School of Engineering, Chiba University, Chiba, 263-8522, Japan
[2]Molecular Chirality Research Centre, Chiba University, Chiba, 263-8522, Japan
[3]Institute for Advanced Academic Research, Chiba University, Chiba, 263-8522, Japan

*E-mail:* [*]asvrao@chiba-u.jp

**Abstract**

We provide a simple analysis based on ray optics and Dirac notation for one and two-dimensional Cosine beams. We then went on to understand the properties of the Bessel beams. For the first time, we report on a generation of three-dimensional needle structures based on interference of one-dimensional Cosine beams. Straightforward mathematical calculations are used to derive the analytical expressions for Cosine beams. The present method of demonstration of Cosine beams may be utilized to understand other structured modes. The Dirac notation-based interference explanation used here can render new researchers to discover an easy way to understand the wave nature of light in fundamental interferometric experiments as well as in advanced-level experiments such as beam engineering technology, imaging, particle manipulation, light sheet microscopy, and light-matter interaction. We also provide an in-depth analysis of similarities among Cosine, Bessel, and Hermite-Gaussian beams.

1. **Introduction**

Transverse structuring of conventional Gaussian-shaped laser beam with phase modulation has produced a variety of structured modes and has seen tremendous applications in fundamental and applied optics. The names of these beams given on the basis of their properties (phase, amplitude, and intensity) follow the mathematical functions. For instance, Laguerre-Gaussian (LG), Hermite-Gaussian (HG) beam, Ince-Gaussian (IG) beam, Bessel beam, Cosine beam, and airy beam can have represented with respective mathematical functions of the Laguerre polynomial, Hermite polynomial, Ince polynomial, Bessel polynomial, cosine function, and airy function [1-7].

Among all structured laser beams, cosine beams have a simple structure and they can be easily synthesized in any optical laboratory with readily available diffractive optical elements like Fresnel biprism, mirror, and beam splitter [8,9]. It has recently been reported that Cosine beams can be generated by using resonant metasurfaces [10]. Some of the works have been carried out on the modulation of Cosine beam characteristics to enhance their user-friendly nature for the scientific community. For instance, the generation and characterization of higher-order Cosine beams [11], truncation of Cos beams [12], creation of apertured Cos beam [13], generation of Cosine-Gauss beam and their propagation through apertured paraxial *ABCD* optical systems [14], and rectification of cosine Gaussian beam [15]. The self-healing and non-diffraction nature [11,16] of Cosine beams make them utilized in fundamental and applied optics. To name a few, various kinds of Cosine beams with and without apodization are used in light-sheet microscopy [17], in the interaction with uniaxial crystal [18], in the interaction of linear and nonlinear media [19,20], in periodic potential optical lattices [21], in the study of spherical particles [22], in optical wireless communication [23], and in plasmonics [24].

On the other side, various kinds of structured laser modes generation, characterization, and their applications mathematically understand by formulating them with ray optics and wave optics. In the understanding of light, ray optics is a very simplified method with multiple limitations, but wave optics have several advantages with containing complex mathematical functions

[25]. Furthermore, there are two other methods that can be utilized to understand the light and which are matrix and Dirac notations [26].

Here, we explore the properties of the Cosine beam in a full-fledged way through Dirac notation and ray optics. We first demonstrate how the individual waves involve in the superposition of optical beams to produce one-dimensional (1D) and two-dimensional (2D) Cosine beams. Further, interpreted their non-diffraction and self-healing properties. The procured results of 1D and 2D Cosine beams motivated us to consider the superposition of two 1D Cosine beams and their result in a series of optical needle arrays. Our analysis extends to Bessel beams and has examined the similarities and differences between Cosine beams, Bessel beams, and HG beams.

In Cosine beams, the transverse intensity modulation is a function of cos. We can produce 1D and 2D Cosine beams by providing cosine intensity modulation in one direction and in two orthogonal directions, respectively. For a trouble-free understanding of Cosine beams, first, we provide a detailed discussion on 1D cosine beams, and then it is extended to 2D Cosine beams.

## 2. One dimensional Cosine beam

1D Cosine beam generation through the superposition of two plane waves of oblique cross propagating and the angle between the two waves is $2\theta$, can be understood through ray optics representation depicted in Fig. 1 (*a*). Let the interference taking place on the *yz*-plane with their bisecting axis be the *z*-axis. Thus, the optical axis of the Cosine beam is given by bisecting axis (*z*-axis) of the two plane waves. The Cosine beam forms only in the overlapping region of the two waves, and its range and width depend on the size and angles of the interfering beams [11]. As shown in Fig. 1(*b*), characteristics of the Cosine beam can be interpreted in terms of the properties of interfering waves. When the interfering two plane waves have polarized perpendicular to the plane of interference, the electric fields of both waves oscillate in the same direction and the interpretation of interference is quite easy. However, when both the plane waves have the same polarization direction but an angle, $\phi$ with respect to the plane of interference, i.e., $\theta$ and $\phi$ directions are orthogonal to each other. Here, whereas $\theta$ is considered from the bisecting axis of the interfering beams (*z*-axis), $\phi$ is from the *x*-axis that is perpendicular to the interference plane. The state of the k-vector and *E*-vector along *x*, *y*, and *z* coordinates for wave-1 and wave-2 are provided in Table 1. The result of counter-propagating *k*-vectors: $k_y$ and $k_y{'}$ produces a standing wave interference along the *y*-axis and the outcome of co-propagating *k*-vectors: $k_z$ and $k_z{'}$ results in the propagation vector of the Cosine beam along the *z*-axis.

Table 1. The propagation vectors and optical field amplitudes of interfering plane waves along *x*, *y*, and *z* directions in the 1D Cosine beam generation.

| Interfering wave | $(k_x, k_y, k_z)$ | $(E_x, E_y, E_z)$ |
|---|---|---|
| Plane wave - 1 | $(0, k\sin\theta, k\cos\theta)$ | $(E\cos\phi, E\sin\phi\cos\theta, E\sin\phi\sin\theta)$ |
| Plane wave - 2 | $(0, -k\sin\theta, k\cos\theta)$ | $(E'\cos\phi, E'\sin\phi\cos\theta, E'\sin\phi\sin\theta)$ |

The generation of a Cosine beam by the interference of two plane waves can be easily understood through bra-ket, $\langle\cdot|\cdot\rangle$ notation. As shown in Fig. 1(*a*), to distinguish interfering waves, we use un-primed coordinates for plane wave - 1 and primed coordinates for plane wave-2. The states of the interfering two plane waves are given by

$$|\alpha\rangle = e^{-iky\sin\theta - ikz\cos\theta}|\beta\rangle, \qquad (1a)$$

$$|\alpha'\rangle = e^{iky\sin\theta - ikz\cos\theta}|\beta'\rangle. \qquad (1b)$$

Here, the three-dimensional polarization states $|\beta\rangle$ and $|\beta'\rangle$ of respective plane waves are given by

$$|\beta\rangle = E\left(\cos\phi|a\rangle + \sin\phi\cos\theta|b\rangle + \sin\phi\sin\theta|c\rangle\right), \qquad (2a)$$

$$|\beta'\rangle = E'\left(\cos\phi|a\rangle + \sin\phi\cos\theta|b\rangle + \sin\phi\sin\theta|c\rangle\right). \qquad (2b)$$

Here, $|a\rangle$, $|b\rangle$, and $|c\rangle$ are polarization states along *x*, *y*, and *z* coordinates respectively. The orthogonality of the polarization states provides the conditions: $\langle a|a\rangle=\langle b|b\rangle=\langle c|c\rangle=1$, and $\langle a|b\rangle=\langle b|c\rangle=\langle c|a\rangle=0$. As a consequence, the inner product of states $|\beta\rangle$ and $|\beta'\rangle$ given by $\langle\beta|\beta\rangle=|E|^2$, $\langle\beta'|\beta'\rangle=|E'|^2$, $\langle\beta|\beta'\rangle=E^*E'$, and $\langle\beta'|\beta\rangle=EE'^*$.

It is a well-known statement that in the interference of two collinearly propagating beams, the interference visibility decreases with increasing the angle between the polarization states of individual beams, and it will be zero when the angle is 90°. However, from the above inner product, we can understand that fringe visibility is independent of polarization for the same initial state of polarization of two interfering beams even when polarization states make some angle with each other. In this scenario, the angular separation between the two polarization states results from the angular separation created between the propagation vectors. It can be inferred from this that the angular separation of the electric field oscillations results from the angular separation of the propagation vectors having no effect on the interference.

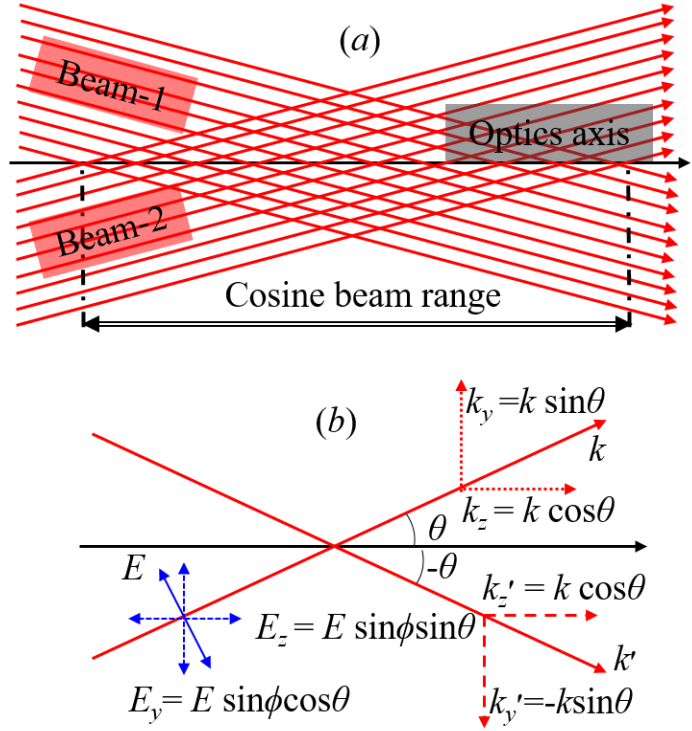

Fig. 1. Ray optics representation of Cosine beam: (*a*) formation of Cosine beam through the interference of two plane waves and (*b*) pictorial interpretation of optical waves' amplitude and propagation vector (dashed arrows are projections).

From Eq. 1, the state of a 1D Cosine beam can be written in terms of interfering beams states $|\alpha\rangle$ and $|\alpha'\rangle$ as

$$|\psi\rangle = e^{-ikz\cos\theta}\left(e^{-iky\sin\theta}|\beta\rangle + e^{iky\sin\theta}|\beta'\rangle\right) \qquad (3)$$

and it is equivalent to the general electric field amplitude expression of a 1D Cosine beam which is given by

$$E(z, y) = E\cos(ky\sin\theta)e^{-ikz\cos\theta}. \qquad (4)$$

The intensity distribution in the 1D Cosine beam given by the inner product of the Dirac state which is provided by Eq. 3 with setting up $E=E'$ as

$$\langle\psi|\psi\rangle = \left(e^{iky\sin\theta}\langle\beta| + e^{-iky\sin\theta}\langle\beta'|\right)\left(e^{-iky\sin\theta}|\beta\rangle + e^{iky\sin\theta}|\beta'\rangle\right)$$
$$\Rightarrow \langle\psi|\psi\rangle = 4I\cos^2(ky\sin\theta) \qquad (5)$$

The 1D Cosine beams can be generated in the experimental laboratory with simple and low-cost experimental configurations as depicted in Fig. 2. A single collimated laser beam can equally divide into two laser beams with a 50:50 beam splitter and further, these two beams can interfere with the aid of a single reflective mirror for the production of 1D Cosine beam. Another way to generate a 1D cosine beam is to illuminate a collimated laser beam on a Fresnel biprism. The biprism with base angle, $\alpha$ divides the circular-shaped laser beam into two semi-circular beams propagating at an angle $\theta = (n-1)\alpha$ with opts axis. Here, the refraction property of biprism enforces the individual waves of the incident collimated laser beam into cross-propagation with respect to the optical axis. The range and size of the Cosine beam depend on the spot size of interfering beams, $w$, and the interfering angle, $\theta$. Even though we use identical input beams in both methods, the cross-sectional intensity distribution of corresponding Cosine beams is different because the method of generation is different. While, the Cosine beam formed in the former technique is a result of the division of amplitude, in the latter method it is due to wave-front division.

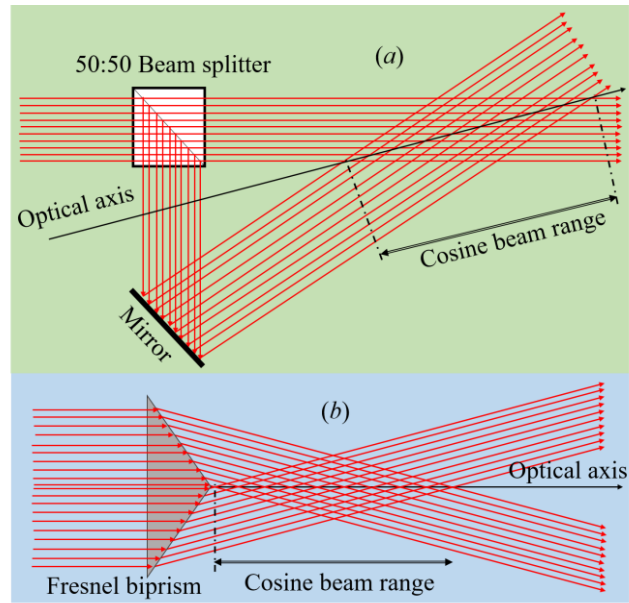

Fig. 2. Schematic experimental diagrams for Cosine beams generation: (a) beam splitter and mirror configuration for Cosine beam generation and (b) Fresnel biprism based Cosine beam generation.

Phase and intensity distributions in the biprism based 1D Cosine beam can be understood through Fig. 3 for plane wave consideration. The linear phase change of biprism in the *y*-direction and corresponding transverse intensity of output 1D Cosine beams are presented in Fig. 3(a) and Fig. 3(b). The transverse phase variation of biprism produced a constant periodic phase throughout the laser beam propagation and as a result, the transverse energy distribution is in the form of the cosine function. For a better understanding of the correlation between the biprism and characteristics of the Cosine beam, we have plotted the line profile of biprisms' phase and Cosine beams' phase, amplitude, and intensity on a single graph as depicted in Fig. 3(*c*).

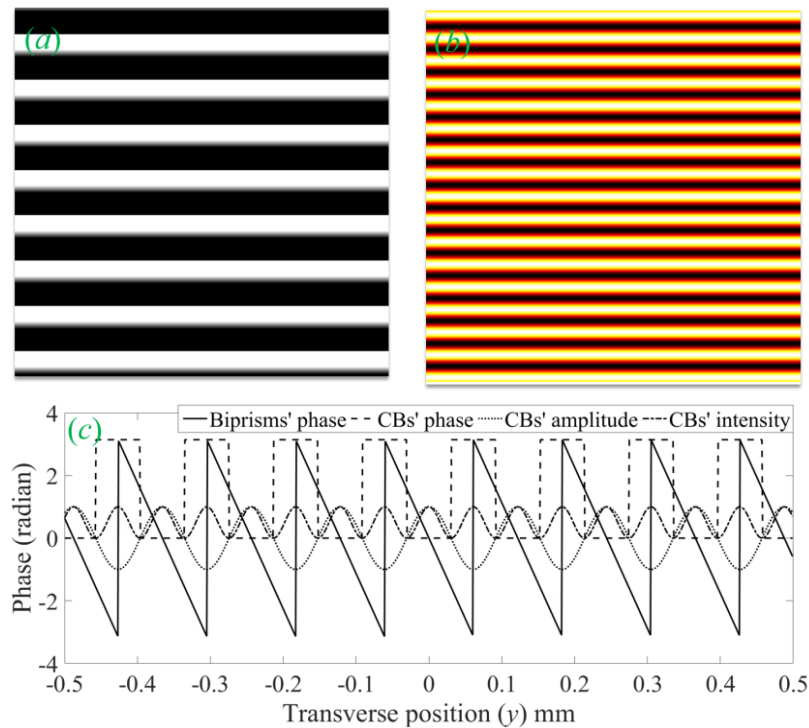

Fig. 3. Properties of Cosine beam generated with Fresnel biprism of base angle $\alpha=1°$ at 640 nm wavelength, (*a*) phase introduced by the biprism to the incident laser beam to produce Cosine beam, (*b*) transverse intensity distribution of Cosine beam, and (*c*) line profiles of biprism and Cosine beam (amplitude and intensity of Cosine beam normalized to one).

## 2.1. Intensity distribution of one dimensional Cosine beam along x, y, and z coordinates

The angular propagation of individual waves in the Cosine beam with respect to its optical axis leads to the possible non-zero electric field components along $x$, $y$, and $z$ directions. With following the condition $E=E'$, the state of the 1D Cosine beam provided by Eq. 3 can be expressed in terms of $x$, $y$, and $z$ states as

$$|\psi_x\rangle = 2E\cos(ky\sin\theta)\cos\phi e^{-ikz\cos\theta}|a\rangle, \quad (6a)$$

$$|\psi_y\rangle = 2E\cos(ky\sin\theta)\sin\phi\cos\theta e^{-ikz\cos\theta}|b\rangle, \quad (6b)$$

$$|\psi_z\rangle = 2E\cos(ky\sin\theta)\sin\phi\sin\theta e^{-ikz\cos\theta}|c\rangle. \quad (6c)$$

The inner product of Eq. 6(a-c) with Eq. 3 produces the electric field intensities along the $x$, $y$, and $z$ directions and are

$$I_x = \langle\psi_x|\psi\rangle = 4I\cos^2(ky\sin\theta)\cos^2\phi, \quad (7a)$$

$$I_y = \langle\psi_y|\psi\rangle = 4I\cos^2(ky\sin\theta)\sin^2\phi\cos^2\theta, \quad (7b)$$

$$I_z = \langle\psi_z|\psi\rangle = 4I\cos^2(ky\sin\theta)\sin^2\phi\sin^2\theta. \quad (7c)$$

The optical field components along the longitudinal and transverse directions depend on the polarization angle, $\phi$, and interfering angle, $2\theta$. For $(\phi, \theta) = (70°, 15°)$, the intensity components of the Cosine beam along the three coordinates, $x$, $y$, and $z$ are given in Fig. 4. For extreme conditions: $(\phi, \theta) = (0°, \theta) \Rightarrow (I_x, I_y, I_z) = (I, 0, 0)$ and $(\phi, \theta) = (90°, \theta) \Rightarrow (I_x, I_y, I_z) = (0, I\cos^2\theta, I\sin^2\theta)$. The $I_z$ is the maximum of $I_z = I / 2$ for $(\phi, \theta) = (90°, 45°)$ and zero for $(\phi, \theta) = (0°, \theta)$ for any arbitrary angles, $(\phi, 0 < \theta < \pi/4)$, $0 < I_z < I / 2$. The longitudinal component, $I_z$ is negligible as compared with the intensity pair $(I_x, I_y)$ for $\theta < 5°$ irrespective of $\phi$ value. Hence, for a few degrees of interfering angle, the effect of $I_z$ can be successfully neglected for any light-matter interaction.

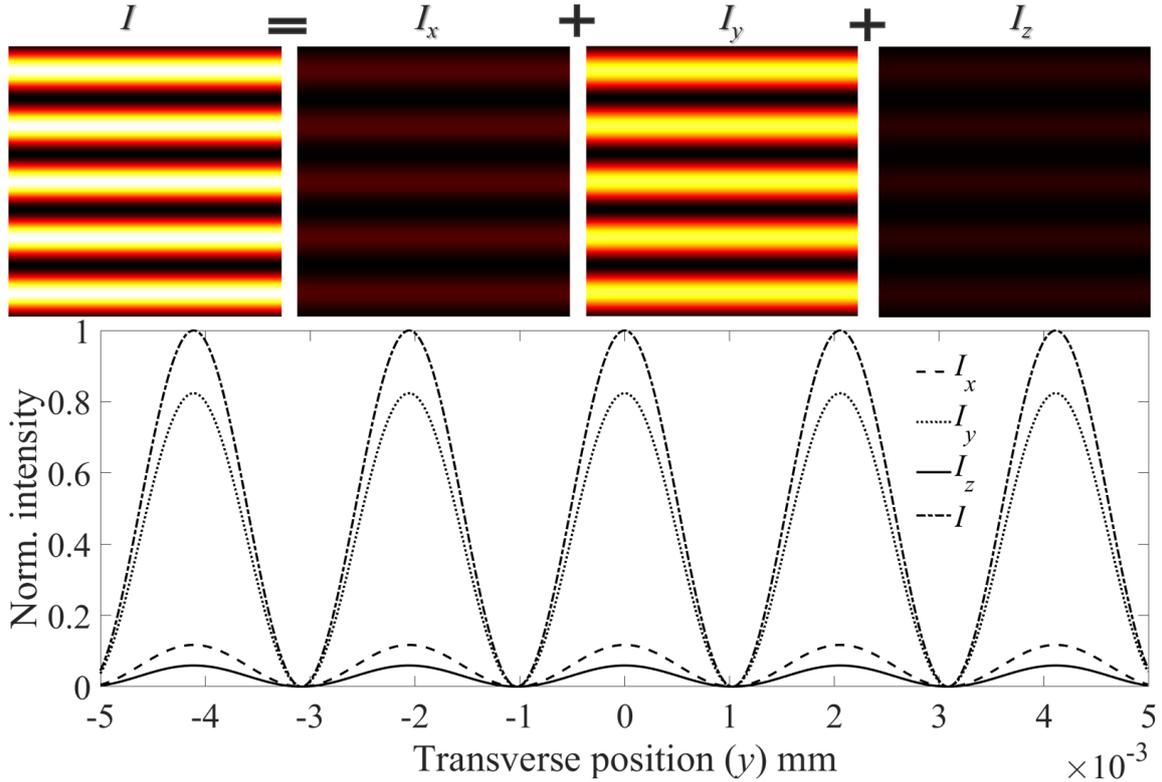

Fig. 4. Cross-sectional intensity components of the Cosine beam in the longitudinal and transverse directions are presented in the first row and corresponding line profiles are plotted in the second row.

## 2.2. Properties of one dimensional Cosine beam

The 1D Cosine beams have one-dimensional non-diffraction and self-healing properties in the plane of interference owing to the individual waves' cross-propagation in an 'X' shape. The energy distribution in the overlapping region of cross-propagation is guided by the interference minima. Hence, without the loss of diffraction property of the wave, the energy distribution in the Cosine beam is free of diffraction. In addition to diffraction-free propagation, 1D Cosine beams have a self-healing property in their plane of incidence. It is also noted that the 1D Cosine beam will not have any self-healing property perpendicular to its plane of incidence due to the absence of cross-propagation of waves. Therefore, 1D Cosine beams can only self-heal for line objects which are placed in the plane of incidence/plane of interference. The self-healing property of a 1D Cosine beam can understand through ray optics as shown in Fig. 5. For line object obstacles, we can see the shadow in a certain region of the Cosine beam, and this shadow depends on the length of the object [11]. The unblocked cross-propagating waves due to the line object produce a Cosine beam before the obstacle and after the shadow region. The fringe visibility [$\eta = (I_{max} - I_{min}) / (I_{max} + I_{min})$] of the Cosine beam is at its maximum value $\eta = 1$ for the case of plane wave superposition.

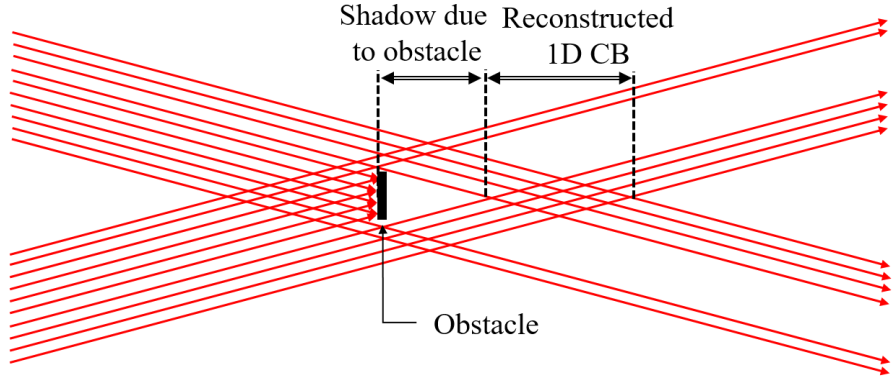

Fig. 5. Visualization of self-healing in 1D Cosine beam through ray representation.

*2. 3. Real Cosine beams*

All mathematical expressions and discussions for the Cosine beam presented above are with due consideration of constant electric field amplitude $E$ provided by the infinite extent of the plane wave. However, this kind of ideal Cosine beam is not realizable in experimental laboratories owing to its infinite extent. In real beams, the transverse intensity distribution is subject to transverse phase distribution, as we discussed in the introduction [1-7]. Hence, the transverse intensity and phase distribution in the Cosine beam is encumbered with the properties of superposed laser beams. It is possible to understand the real Cosine beams through the truncation of Cosine beams. Depending on the interfering beams, we can apodize the Cosine beam, and we can create various types of Cosine beams like Cosine-Gauss / Cos-Gauss (CG) beam [19], Cosine- Hermite-Gauss (CHG) beam [11], Super-Gaussian-Cosine (SGC) beam [7], etc. As shown in Fig. 6, the superposition of two Gaussian beams and superposition of two HG beams are utilized to generate the respective CG beam and CHG beam based on the method given in Fig. 2(a). The optical field amplitude formed as a result of superposition can be written in terms of individual field amplitudes as $E = E(x, y+y_0, z) + E'(x, y-y_0, z)$ and $y_0 = z \tan\theta$. In a similar fashion, we can generate the Cosine-Ince-Gauss beam and Cosine-Laguerre-Gauss beam. The conspicuous non-uniform intensity in the real laser modes produces Cosine beams with position-dependent fringe visibility. Therefore, the condition $E=E'$ is no longer valid in real beams and the position-dependent fringe visibility can have it's all values, i.e., $0 \leq \eta (r, z) \leq 1$. The fringe visibility is equal to one near close to the optical axis, and it decreases with increasing the transverse position of the beam. The position-dependent fringe visibility in the real beams can be seen unquestionably in Fig. 6.

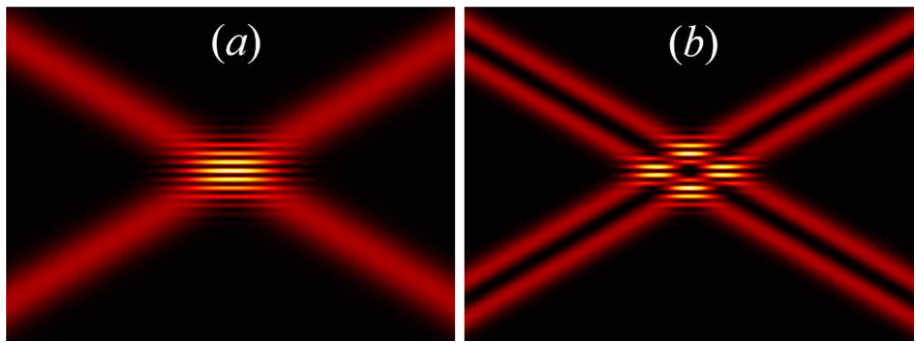

Fig. 6. 1D Cosine beam formed as a result of two beams cross-interference provided by the experimental configuration shown in Fig. 2(a): (*a*) Cosine-Gaussian beam in the presence of superposition of two Gaussian beams, and (*b*) Cosine Hermite-Gauss beam under superposition of two $HG_{01}$ modes (here, $\theta = 5°$).

## 3. Two dimensional Cosine beam

2D Cosine beam in bra-ket notation is a mixed state and it is formed by the product of two pure states of 1D Cosine beams. The two 1D Cosine beams must have the same optical axis with the orthogonal plane of incidence. The 2D Cosine beam state is given by $|\Psi\rangle = |\psi_x\rangle |\psi_y\rangle$. Here, the 1D Cosine beam with the plane of incidence is *xz*-plane as

$$|\psi_x\rangle = e^{-ikz\cos\theta_x}\left(e^{-ikx\sin\theta_x}|\beta_x\rangle + e^{ikx\sin\theta_x}|\beta'_x\rangle\right) \quad (8a)$$

and the state of the 1D Cosine beam with the plane of incidence is *yz*-plane as

$$|\psi_y\rangle = e^{-ikz\cos\theta_y}\left(e^{-iky\sin\theta_y}|\beta_y\rangle + e^{iky\sin\theta_y}|\beta'_y\rangle\right) \quad (8b)$$

The 2D Cosine beam state $|\Psi\rangle$ in Dirac notation is equivalent to the general optical field amplitude expression of the 2D Cosine beam given by [18]

$$E(z,r) = E\cos(kx\sin\theta_x)e^{-ikz\cos\theta_x}\cos(ky\sin\theta_y)e^{-ikz\cos\theta_y}. \quad (9)$$

Here, states $|\beta_i\rangle$ and $|\beta'_i\rangle$ are functions of $\theta_x, \phi$ in the *xz*-plane and are functions of $\theta_y, \phi$ in the *yz*-plane (detailed information can be found in section A of the supplementary file). The intensity distribution in the 2D Cosine beam is given by the inner product of its Dirac state. Each ket state inner product with its corresponding bra state is given by

$$\langle\Psi|\Psi\rangle = \langle\psi_x|\psi_x\rangle\langle\psi_y|\psi_y\rangle. \quad (10)$$

From Eq. 3 and Eq. 10, we can obtain the Dirac product as

$$\langle\Psi|\Psi\rangle = 8I^2\cos^2(kx\sin\theta_x)\cos^2(ky\sin\theta_y). \quad (11)$$

It is noted that $|\Psi\rangle$ can further written as $|\Psi\rangle = |\psi_x \psi_y\rangle$ instead of $|\Psi\rangle = |\psi_x\rangle|\psi_y\rangle$. However, the Eq. $|\Psi\rangle = |\psi_x \psi_y\rangle$ is more complex than the $|\Psi\rangle = |\psi_x\rangle|\psi_y\rangle$, however, both the Eqs. provide the same result as shown in Fig. 7. The derivation for Eq. $|\Psi\rangle = |\psi_x \psi_y\rangle$ can be found in section A of the supplementary file.

2D Cosine beams have non-diffraction and self-healing properties similar to the 1D Cosine beam. These properties are more effective for rectangular objects. The cosine modulations along the *x* and *y* directions of the 2D Cosine beam create the needle array structures [27] in it. The needle array with their self-healing and non-diffraction properties can have used in the on-axis trapping of multiple particles in multiple parallel planes simultaneously [28], parallel guiding of multiple particles [29], interstitial imaging of solid tissues and organs through optical coherence tomography imaging needle [30] circular dichroism imaging of sparse sub-diffraction objects [31]. The longitudinal axis of the needle array is parallel to the optical axis of the Cosine beam. The intensity and phase of the 2D CG beam produced in the Gaussian host are presented in Fig. 7. Here, we used $\theta_x = \theta_y = 5°$ to produce square shape needles. We can also generate rectangular needles under $\theta_x \neq \theta_y$ condition sans any on-axis intensity modulation.

The 2D CG beam can be experimentally realized with SLM or DMD by projecting the phase hologram of Fig. 7(*a*) on them. As we discussed in the 1D Cosine beam, we can also produce complex structures in the 2D Cosine beam by simply changing the host beam.

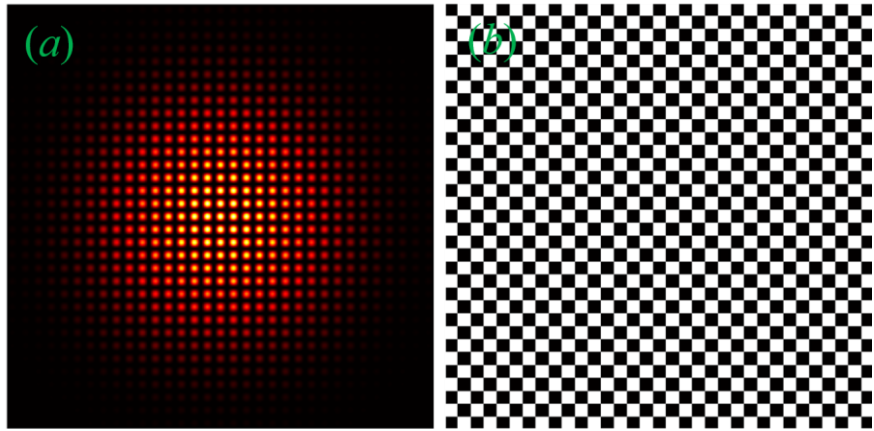

Fig. 7. (*a*) Intensity distribution and (*b*) phase of 2D Cosine beam (here, $\theta_x = \theta_y = 5°$).

*3. 1. Intensity distribution of 2D Cosine beam along x, y, and z coordinates*

If the peak intensity of 1D and 2D Cosine beams is considered as $I_0$ then the z-component intensity of 1D Cosine beam with the plane of interference is $xz$- plane ($I_1$), 1D Cosine beam with the plane of interference is $yz$- plane ($I_2$), and 2D Cosine beam formed by the interference of first two 1D Cosine beams ($I_3$) given by Eq. 13.

$$I_1 = I_0 \cos^2\phi \sin^2\theta, \quad (12a)$$

$$I_2 = I_0 \sin^2\phi \sin^2\theta, \quad (12b)$$

$$I_3 = \frac{1}{2}(I_1 + I_1) = \frac{1}{2} I_0 \sin^2\theta, \quad (12c)$$

For simplicity, here, we considered the condition $\theta_x = \theta_y = \theta$ in the 2D Cosine beam. The expressions provided in Eq. 12 are procured from Fig. 1(b). While the z-component of the optical field is depending on angles, $\phi$ and $\theta$ in 1D Cosine beam, it is independent of $\phi$ in 2D Cosine beam attributed to cross-propagating waves in both $xz$ and $yz$ planes. However, $I_z$ can be controlled with $\theta$. The maximum of $I_z$ in a 2D Cosine beam is less than the corresponding value in a 1D Cosine beam. For example, at $\theta = 45°$ the max($I_z$) = $I/4$ in the 2D Cosine beam and max($I_z$) = $I/2$ for 1D Cosine beam.

## 4. Superposition of one-dimensional Cosine beams with same polarization

1D Cosine beam can be understood through the superposition of two plane waves in the cross propagation, as we discussed in section 2. In the same vein, we can understand the superposition of two 1D Cosine beams of $|\psi_x\rangle$ and $|\psi_y\rangle$ by the consideration of four-wave interference. Now consider angularly propagating four identical waves emanating from a single laser source. Two are in the $xz$-plane and the other two are in the $yz$-plane, i.e., planes of incidence are orthogonal with each other. Let the four waves be linearly polarized at an angle $\phi$ with respect to the $x$-axis. Let the angle created by these waves in the $xz$-plane and $yz$-plane with reference to the $z$-axis be $\theta_x$, and $\theta_y$ respectively. Then the state of the k-vector and $E$-vector of four waves are present along the directions of $x$, $y$, and $z$ coordinates and are given in Table 2. The two vectors have the form of $(k_x, k_y, k_z) = (k\sin\theta_x, k\sin\theta_y, k\cos\theta_x + k\cos\theta_y)$ and $(E_x, E_y, E_z) = (E\cos\phi\cos\theta_x, E\sin\phi\cos\theta_y, E\cos\phi\sin\theta_x + E\sin\phi\sin\theta_y)$.

Table 2. The propagation vectors and optical field amplitudes of interfering beams along $x$, $y$, and $z$ directions in four wave-interference.

| Interfering wave | ($k_x$, $k_y$, $k_z$) | ($E_x$, $E_y$, $E_z$) |
|---|---|---|
| Plane wave - $x_1$ | ($k\sin\theta_x$, 0, $k\cos\theta_x$) | ($E\cos\phi\cos\theta_x$, $E\sin\phi$, $E\cos\phi\sin\theta_x$) |
| Plane wave - $x_2$ | ($-k\sin\theta_x$, 0, $k\cos\theta_x$) | ($E'\cos\phi\cos\theta_x$, $E'\sin\phi$, $E'\cos\phi\sin\theta_x$) |
| Plane wave - $y_1$ | (0, $k\sin\theta_y$, $k\cos\theta_y$) | ($E\cos\phi$, $E\sin\phi\cos\theta_y$, $E\sin\phi\sin\theta_y$) |
| Plane wave - $y_2$ | (0, $-k\sin\theta_y$, $k\cos\theta_y$) | ($E'\cos\phi$, $E'\sin\phi\cos\theta_y$, $E'\sin\phi\sin\theta_y$) |

The states of four waves provided through their propagation parameters are as follows

$$|\alpha_{x1}\rangle = e^{-ikx\sin\theta_x - ikz\cos\theta_x}|\beta_x\rangle, \quad (13a)$$

$$|\alpha_{x2}\rangle = e^{ikx\sin\theta_x - ikz\cos\theta_x}|\beta'_x\rangle, \quad (13b)$$

$$|\alpha_{y1}\rangle = e^{-iky\sin\theta_y - ikz\cos\theta_y}|\beta_y\rangle, \quad (13c)$$

$$|\alpha_{y2}\rangle = e^{iky\sin\theta_y - ikz\cos\theta_y}|\beta'_y\rangle. \quad (13d)$$

Here, the three-dimensional polarization states $|\beta\rangle$ and $|\beta'\rangle$ of respective waves are the same as given in Eq. 2 (detailed expressions can find in the supplementary file as Eq. $S_2$). The state of superposition can be written in terms of the state of interfering waves as

$$|\psi\rangle = e^{-ikz\cos\theta_x}\left(e^{-ikx\sin\theta_x}|\beta_x\rangle + e^{ikx\sin\theta_x}|\beta'_x\rangle\right) + e^{-ikz\cos\theta_y}\left(e^{-iky\sin\theta_y}|\beta_y\rangle + e^{iky\sin\theta_y}|\beta'_y\rangle\right) \quad (14)$$

The intensity distribution of the superposition of two 1D Cosine beams is given by the inner product as

$$\langle\psi|\psi\rangle = \left[e^{ikz\cos\theta_x}\left(e^{ikx\sin\theta_x}\langle\beta_x| + e^{-ikx\sin\theta_x}\langle\beta'_x|\right) + e^{ikz\cos\theta_y}\left(e^{iky\sin\theta_y}\langle\beta_y| + e^{-iky\sin\theta_y}\langle\beta'_y|\right)\right]$$
$$\left[e^{-ikz\cos\theta_x}\left(e^{-ikx\sin\theta_x}|\beta_x\rangle + e^{ikx\sin\theta_x}|\beta'_x\rangle\right) + e^{-ikz\cos\theta_y}\left(e^{-iky\sin\theta_y}|\beta_y\rangle + e^{iky\sin\theta_y}|\beta'_y\rangle\right)\right]$$

After simple theoretical calculations, we deduce a simple expression for the intensity of the interference as

$$\langle\psi|\psi\rangle = 4I\left[\cos^2(kx\sin\theta_x) + \cos^2(ky\sin\theta_y)\right] + 8I\cos\left[kz(\cos\theta_x - \cos\theta_y)\right]\cos(kx\sin\theta_x)\cos(ky\sin\theta_y)$$
$$\left(\cos^2\phi\cos\theta_x + \sin^2\phi\cos\theta_y + \cos\phi\sin\phi\sin\theta_x\sin\theta_y\right) \qquad (15)$$

The full derivation of Eq. 15 can be found in section B of the supplementary file.

*4.1. The intensity of superposition of one-dimensional Cosine beams long x, y, and z directions*

By following the condition of $E=E'$, the state vector of interference in the $x$, $y$, and $z$ direction are given by

$$|\psi_x\rangle = \left[e^{-ikz\cos\theta_x} 2E\cos(kx\sin\theta_x)\cos\phi\cos\theta_x + e^{-ikz\cos\theta_y} 2E\cos(ky\sin\theta_y)\cos\phi\right]|a\rangle, \qquad (16a)$$

$$|\psi_y\rangle = \left[e^{-ikz\cos\theta_x} 2E\cos(kx\sin\theta_x)\sin\phi + e^{-ikz\cos\theta_y} 2E\cos(ky\sin\theta_y)\sin\phi\cos\theta_y\right]|b\rangle, \qquad (16b)$$

$$|\psi_z\rangle = \left[e^{-ikz\cos\theta_x} 2E\cos(kx\sin\theta_x)\cos\phi\sin\theta_x + e^{-ikz\cos\theta_y} 2E\cos(ky\sin\theta_y)\sin\phi\sin\theta_y\right]|c\rangle. \qquad (16c)$$

By the product of Eq. 14 and Eq. 16, we can obtain the intensities along the $x$, $y$, and $z$ directions are

$$I_x = \langle\psi_x|\psi\rangle = 4I\cos^2(kx\sin\theta_x)\cos^2\phi\cos^2\theta_x + 4I\cos^2(ky\sin\theta_y)\cos^2\phi$$
$$+ 8I\cos(kx\sin\theta_x)\cos(ky\sin\theta_y)\cos^2\phi\cos\theta_x\cos\left[kz(\cos\theta_x - \cos\theta_y)\right], \qquad (17a)$$

$$I_y = \langle\psi_y|\psi\rangle = 4I\cos^2(kx\sin\theta_x)\sin^2\phi + 4I\cos^2(ky\sin\theta_y)\sin^2\phi\cos^2\theta_y$$
$$+ 8I\cos(kx\sin\theta_x)\cos(ky\sin\theta_y)\sin^2\phi\cos\theta_y\cos\left[kz(\cos\theta_x - \cos\theta_y)\right], \qquad (17b)$$

$$I_z = \langle\psi_z|\psi\rangle = 4I\cos^2(kx\sin\theta_x)\cos^2\phi\sin^2\theta_x + 4I\cos^2(ky\sin\theta_y)\sin^2\phi\sin^2\theta_y$$
$$+ 8I\cos(kx\sin\theta_x)\cos(ky\sin\theta_y)\sin\phi\cos\phi\sin\theta_x\sin\theta_y\cos\left[kz(\cos\theta_x - \cos\theta_y)\right]. \qquad (17c)$$

From the above expressions, we can understand that the total intensity as well as intensities along the $x$, $y$, and $z$ directions of interference depends on angles: $\phi$, $\theta_x$, and $\theta_y$. The optical field component along the $z$-direction depends on the angles, $\theta_x$, and $\theta_y$ in a similar fashion to 1D Cosine beams. For a clear systematic understanding of the interfering intensity distribution, the total intensity of the interference as a result of two 1D Cosine beams is normalized to one. Hence the fractional optical intensity of each 1D Cosine beam is ½. The calculated fractional intensities of two 1D Cosine beams for $\phi = 0°$, 90°, and 45° are given in Table 3. From this table we can calculate the fractional intensity distribution of the interference pattern along the $x$, $y$, and $z$ directions, and corresponding intensity distributions are shown in Fig. 8 for $\theta_x = \theta_y = 45°$. For $\phi = 0°$, 1D Cosine beam in the $xz$-plane has cosine intensity distribution along $x$ - direction for electric field components $|\psi_x\rangle$ and $|\psi_z\rangle$ with $|\psi_y\rangle = 0$. For the same case, the 1D Cosine beam in the y$z$-plane has cosine intensity distribution along $y$ - the direction for electric field component $|\psi_x\rangle$ with $|\psi_y\rangle = |\psi_z\rangle = 0$. The unequal intensities of the two 1D Cosine beams in the $x$ - direction produced elliptical array spots with its major axis along the $x$-direction. This effect further produces a mesh shape in the total intensity. A similar effect can also be seen for $\phi = 90°$. We can see the effects produced in the interference for $\phi = 0°$, and $\phi = 90°$ cases simultaneously with an equal contribution for the consideration of $\phi = 45°$. For this condition, the total intensity is in the form of a 2D Cosine beam with its direction in the diagonal. In the present context, the third case is nothing but a superposition of the first two cases. From this, we can conclude that the spatial distributions of intensities: $I$, $I_x$, $I_y$, and $I_z$ can be controlled with $\phi$, $\theta_x$, and $\theta_y$.

Table. 3. Fractional intensities along the $x$, $y$, and $z$ directions for $\theta_x = \theta_y = 45°$ for three angle of polarizations.

| Angle $\phi$ | Type of 1D Cosine beam | Intensities | | |
|---|---|---|---|---|
| | | $I_x$ | $I_y$ | $I_z$ |
| 0° | 1D Cosine beam in $xz$-plane | 1/4 | 0 | 1/4 |
| | 1D Cosine beam in $yz$-plane | 1/2 | 0 | 0 |
| 90° | 1D Cosine beam in $xz$-plane | 0 | 1/2 | 0 |
| | 1D Cosine beam in $yz$-plane | 0 | 1/4 | ¼ |
| 45° | 1D Cosine beam in $xz$-plane | 1/8 | 1/4 | 1/8 |
| | 1D Cosine beam in $yz$-plane | 1/4 | 1/8 | 1/8 |

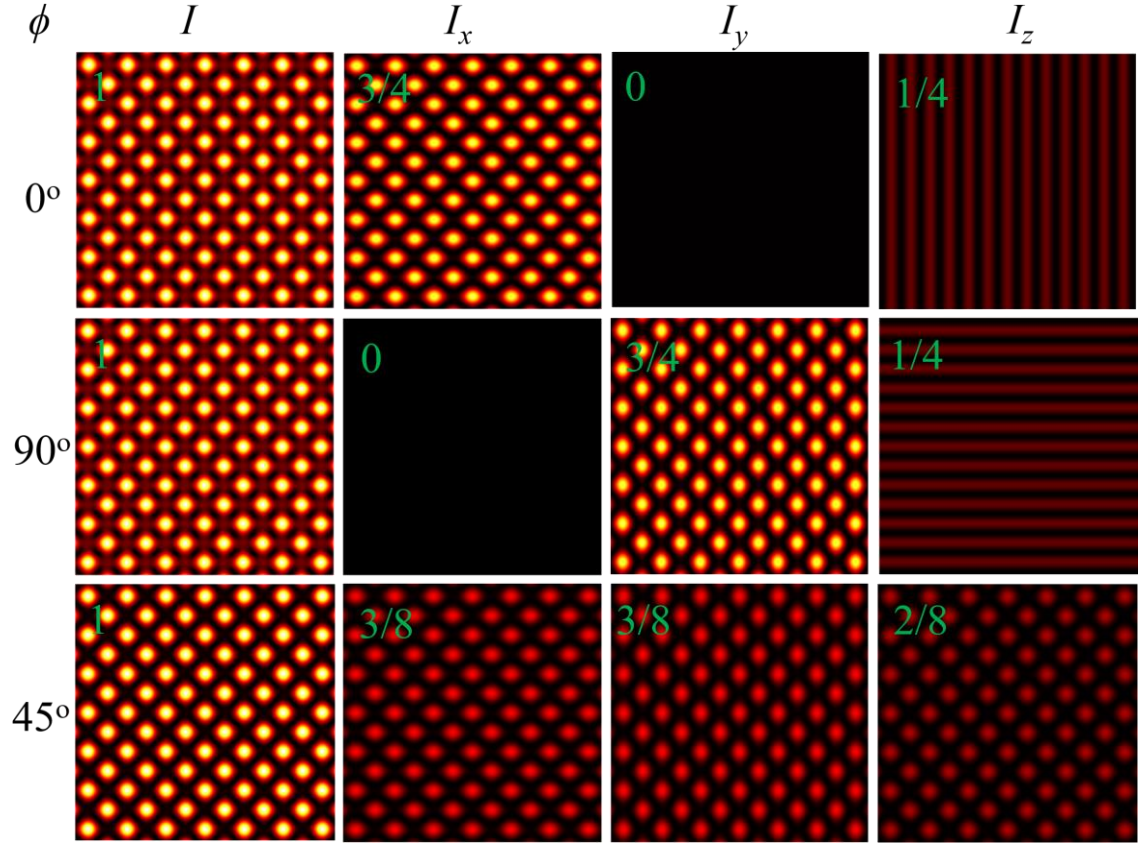

Fig. 8. The transverse intensity distribution of superposed 1D Cosine beams in the *xy*-plane for $\phi = 0°$, $90°$, and $45°$. Fractional intensities of interference along *x*, *y*, and *z* directions are provided in Green font in the respective images (here, $\theta_x = \theta_y = 45°$).

In addition to different kinds of 2D structures creation in the superposition of two 1D Cosine beams under $\theta_x = \theta_y$ condition, we can also produce 3D structures by the consideration of $\theta_x \neq \theta_y$. In Eq.15, we can perceive that the intensity distribution of superposition modes depends on *z*, $\theta_x$, and $\theta_y$ as $\cos[kz(\cos\theta_x - \cos\theta_y)]$. From this, we can understand that we can produce periodic changes in the superposition beam along the beam propagation direction with $\theta_x \neq \theta_y$. Thus, the superposition of 1D Cosine beams can produce the same structure in the regular interval along the propagation without the aid of any diffractive optical elements and this property is called as Talbot effect or self-imaging [32,33]. From the nature of self-imaging, we can understand that the superposition of 1D Cosine beams can generate a series of optical needle arrays with due consideration of unequal angles of $\theta_x$, and $\theta_y$. In Fig. 9, we have shown a series of optical needle arrays for $\theta_x - \theta_y = 10°$. The unequal $\theta_x$ and $\theta_y$ give rise to an elliptical shape needle and this ellipticity nature can augment with increasing $\theta_x - \theta_y$. The tails of the needles overlap with each other along the propagation. The self-healing and non-diffraction nature of optical needles originated from their parent 1D Cosine beams owing to their inherited property. The needle length and their visibility decrease with increasing the angular difference $\theta_x - \theta_y$ and this effect can be seen in the line profile plot of Fig. 9 along the needle propagation. This is the first report on the demonstration of a series of optical needle arrays. The control tunable length and shape of needle arrays owing to their non-diffraction and self-healing can enrich their applications in material science. The interference of two 1D Cosine beams can experimentally realize in a cost-effective way by using two orthogonally rotated Fresnel biprisms in tandem in a single laser beam.

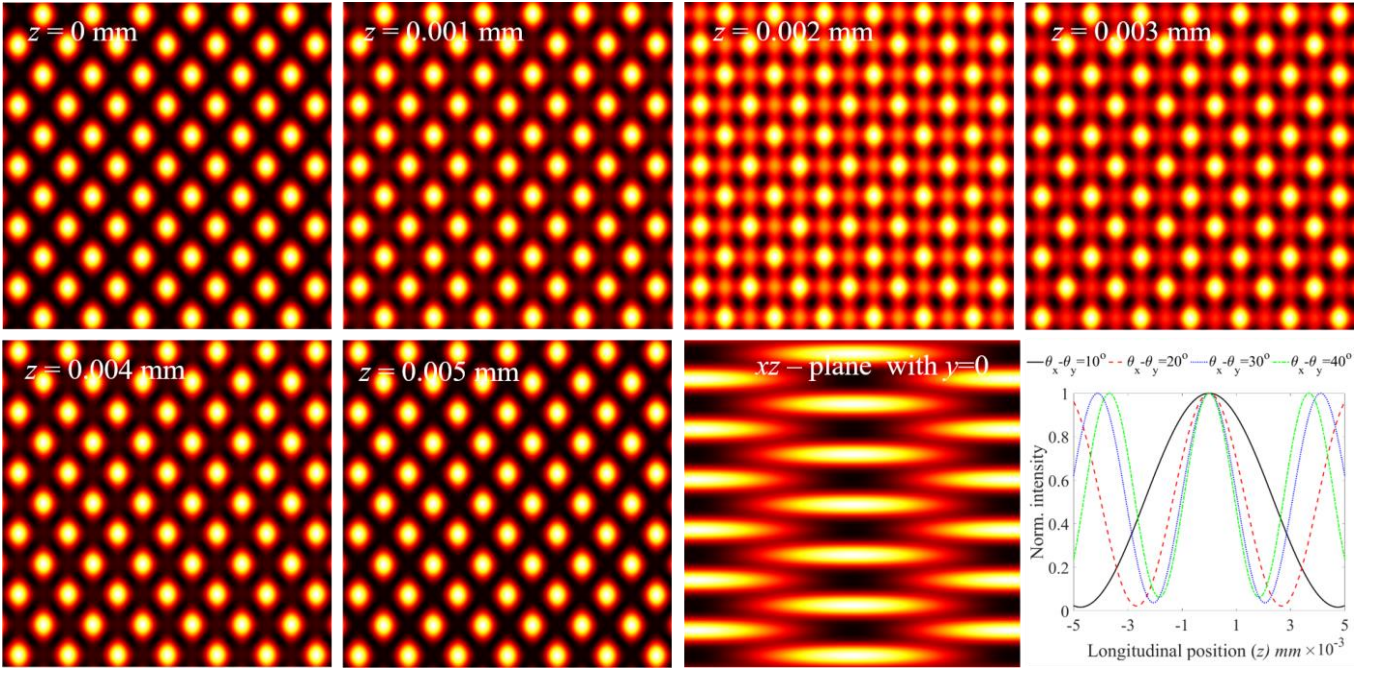

Fig. 9. An optical needle array in series created by two one-dimensional plane wave Cosine beams in the presence of $\theta_x - \theta_y = 10°$. first row and second row first two images are transverse intensity distributions of optical needle array for various longitudinal positions. The last two-dimensional image is a longitudinal cross-section of the needle beam in the $xz$-plane with $y=0$. The dependency of periodicity or size of optical needles on $\theta_x - \theta_y$ is shown in the line plot.

## 5. Superposition of two orthogonally polarized one dimensional Cosine-Gauss beams

Let us consider two pairs of collimated Gaussian beams interfering in the cross-propagation with respect to the $z$-axis. The first pair of beams interfere in the $xz$-plane with their polarization in the $xz$-plane to produce a 1D CG beam of $|\psi_x\rangle$ and the second pair of beams interfere in the $yz$-plane with their polarization in the $yz$-plane to create 1D CG beam of $|\psi_y\rangle$ in $yz$-plane. The resultant superposed state $|\psi\rangle$ in the form of a vector matrix is given by

$$|\psi\rangle = |\psi_x\rangle|\leftarrow\rangle + |\psi_y\rangle|\uparrow\rangle \qquad (18)$$

and horizontal and vertical polarization states are given by

$$|\leftarrow\rangle = \begin{bmatrix} 1 \\ 0 \end{bmatrix} \text{ and } |\uparrow\rangle = \begin{bmatrix} 0 \\ 1 \end{bmatrix}. \qquad (19)$$

Here we consider the angle between the beams in $xz$-plane ($\theta_x$) and $yz$-plane ($\theta_y$) to 5° ($\theta_x = \theta_y = 5°$). Thus, the $z$ component of the electric field can be successfully neglected for simplicity. The intensity and phase distributions of interference are in mesh shape as shown in Fig. 10 (*a*) and Fig. 10(*e*). The single polarization state of this 2D Cosine beam can be understood in the presence of a linear polarizer provided a matrix operator

$$P(\theta_p) = \begin{bmatrix} \cos^2\theta_p & \cos\theta_p \sin\theta_p \\ \cos\theta_p \sin\theta_p & \sin^2\theta_p \end{bmatrix}. \qquad (20)$$

Here, $\theta_p$ is the angle of the transmission axis of the linear polarizer with reference to the $x$-axis. In the presence of a polarizer, the transverse intensity and phase of the superposition state can be controlled with its polarization angle, $\theta_p$. For instance, P(0°) $|\psi\rangle=|\psi_x\rangle|\leftarrow\rangle$, P(90°) $|\psi\rangle=|\psi_y\rangle|\uparrow\rangle$, and P(45°) $|\psi\rangle=|\psi_x\rangle|\nearrow\rangle+|\psi_y\rangle|\nearrow\rangle$ or P(135°) $|\psi\rangle=|\psi_x\rangle|\nwarrow\rangle+|\psi_y\rangle|\nwarrow\rangle$ bring outs horizontal 1D CG beam and vertical 1D CG beam and diagonal or anti-diagonal 2D CG beam respectively. The intensity distributions of diagonal or anti-diagonal 2D CG beams can be further modified by controlling the phase delay between the superposed 1D CG beams or by independently changing the angles, $\theta_x$, and $\theta_y$. Detailed mathematical expressions for Superposition of two orthogonally polarized one dimensional Cosine beams derived in in section C of the supplementary file.

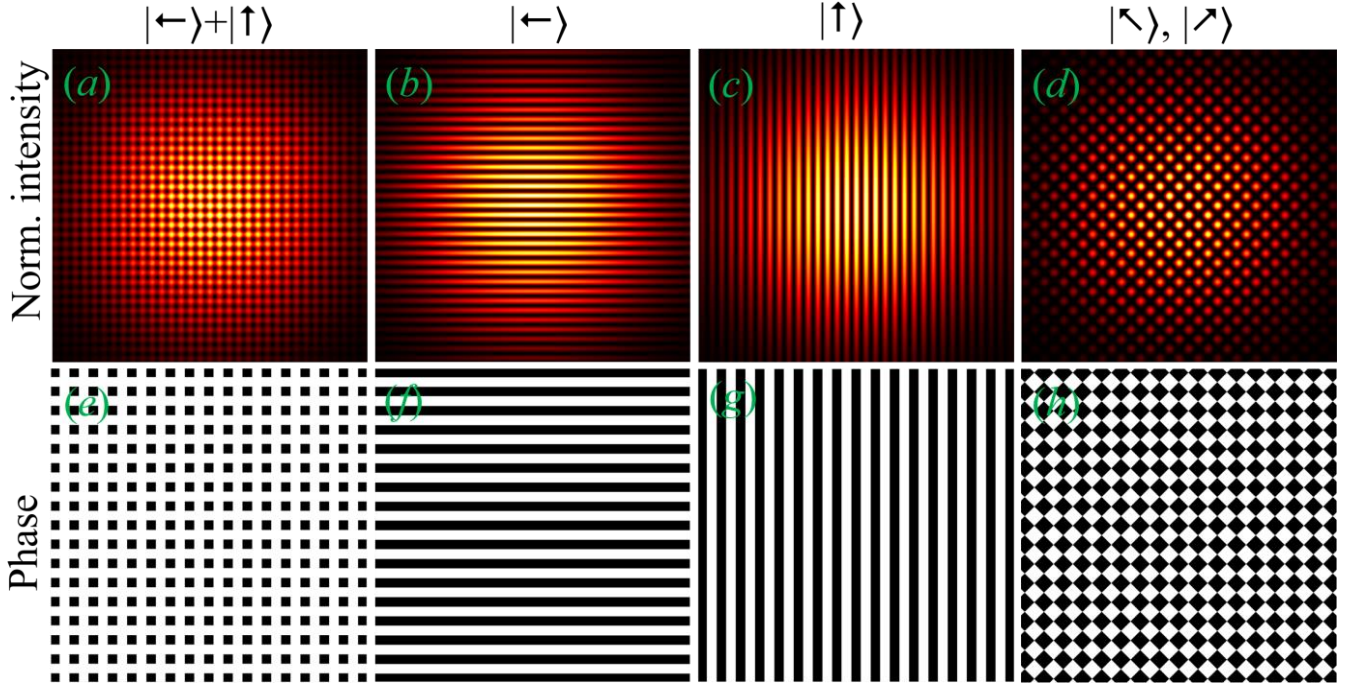

Fig. 10. Generation of various kinds Cosine-Gauss beams by the superposition of two one-dimensional Cosine-Gauss beams with orthogonally polarized. The first and second rows are the intensity and phase of the Cosine-Gauss beam respectively for different states of polarization. Arrows in the ket notation indicate the polarization state of the corresponding Cosine-Gauss beam (here, $\theta_x = \theta_y = 5º$).

## 6. Bessel beam

As we discussed in previous sections, the 1D Cosine beam and 2D Cosine beam are formed as a consequence of the superposition of two and four-plane waves. To create a Bessel beam, we must need an infinite number of plane waves angularly propagating with respect to the $z$-axis. Eventually, the phase and intensity became the Bessel function instead of the cosine function as we depicted in Fig. 11. The Bessel beam can be generated by the superposition of 1D Cosine beams and can be proved with the below two mathematical expressions. The Bessel function is equal to the Sinc function [34]

$$J_0(r) = \frac{\sin r}{r}. \qquad (21)$$

For $|r|<1$, the Sinc function can be written in terms of the cos function as [35]

$$\frac{\sin r}{r} = \prod_{m=1}^{\infty} \cos(r/2^n). \qquad (22)$$

From the above two expressions, zeroth-order Bessel function can be written in terms of the cos function as

$$J_0(r) = \prod_{m=1}^{\infty} \cos(r/2^n). \qquad (23)$$

Hence, to produce a Bessel beam, we just need a product of an infinite number of 1D Cosine beams. In mathematical analysis, it looks very elusive, however, we can very easily achieve it in experiments through a single axicon.

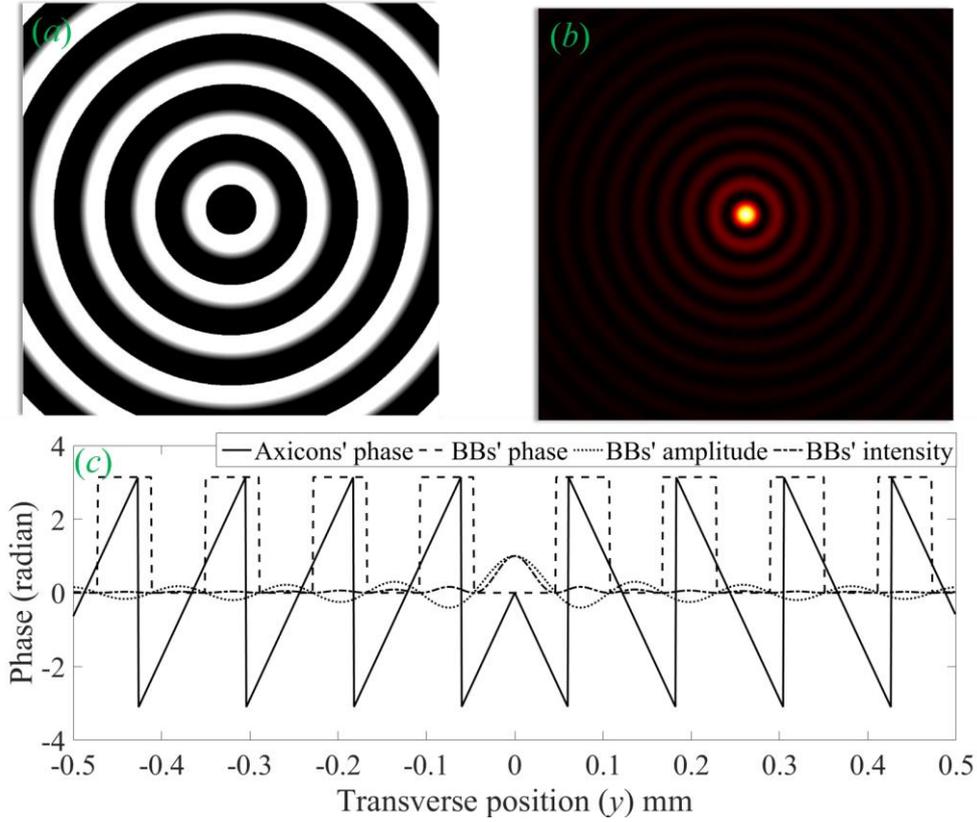

Fig. 11. Properties of Bessel beam generated with axicon of base angle $\alpha=1°$ at 640 nm wavelength, (a) phase introduced by axicon to the incident laser beam for Bessel beam generation, (b) transverse intensity distribution of Bessel beam, and (c) line profiles of axicon and Cosine beam (amplitude and intensity of Cosine beam are normalized to one).

**7. Comparative analysis between Hermite-Gaussian beam, Cosine beam, and Bessel beam**

The both HG beam and the Cosine beam have separable solutions in $x$ and $y$. Hence they can be written as the product of a factor in $x$ and a factor in $y$. 1D and 2D cosine modes in the Cosine beam are similar to the 1D and 2D HG modes in the HG beam [4]. Ideal Bessel and Cosine beams are not square-integrable owing to their infinite energy distribution [3]. However real Bessel and Cosine beams are square-integrable due to their finite size and energy.

Non-uniform energy distribution present in HG beams with corner lobes have maximum peak intensity. However, an ideal Cosine beam has uniform energy distribution in its cross-section with all lobes having the same peak intensity. In a similar way, optical energy in all rings of the ideal Bessel beam is equal but the peak intensity of rings decreases with increasing their radius. In the experimental realization of the Cosine beams and Bessel beams, the peak intensity and energy in each lob decrease while moving away from the beam axis [28].

Experimentally realized Cosine and Bessel beams as a consequence of interference and are not Eigenmodes. Hence, their shape can preserve only for finite propagation. The HG modes are Eigenmodes of the paraxial wave equation and have a stable mode structure with propagation independence. While the Cosine beam has no mode number, HG mode has a mode number and is given by $N=m+n$ [4]. Here, $m$ and $n$ are mode orders along the $x$ and $y$ directions. Both the Cosine and HG beams have no helical wave-front. Bessel beam quantitatively understands with its mode number $l$ and it is also called orbital angular momentum. Bessel beams can have a helical wave-front.

While Cosine and Bessel beams have a self-healing and non-diffraction nature, HG beam doesn't have these properties. 1D and 2D Cosine beams have non-diffraction and self-healing in rectangular symmetry. Therefore, these properties are effective for rectangular objects whose shape is similar to the lobes in the Cosine beam. The Bessel beam has non-diffraction and self-healing [28] in cylindrical symmetry and these properties are more effective for circular objects. Both of these beams have these unique properties attributed to the interference of cross-propagating waves along the propagation. Under paraxial propagation, the HG beam has a single propagation vector $k=2\pi/\lambda$ and it is along its optical axis. However, the 1D Cosine beam has two $k$ vectors, the 2D Cosine beam has four $k$ vectors, and the Bessel beam has an infinite number of $k$ vectors. Their propagation vectors make an angle with the optical axis.

1D Cosine beam can have a $z$-dependent intensity component, $I_z$ only if it is polarized in the plane of interference. 2D Cosine beam, Bessel beam, and focused HG beam can always have $I_z$ irrespective of their polarization direction. If the 1D Cosine beam has its polarization in its plane of interference, then the maximum of $I_z$ in the 2D Cosine beam is less than the corresponding value in the 1D Cosine beam. For any arbitrary linear polarization, the z-component of the optical field in the Bessel beam follows the same trend as in the 2D Cosine beam.

As we discussed in section 4, the interference of 1D Cosine beams can produce a non-diffracting and self-healing series of optical needle arrays. On the other side interference of Bessel beams can also generate non-diffracting and self-healing series of optical bottles, three-dimensional optical potentials [36-38].

In addition to scalar modes, we can also generate spatially varying polarization in Cosine beams and Bessel beams, and HG beams. While the polarization changes in the beam cross-section are in radial and azimuthal directions with cylindrical symmetry in the Bessel beams [39], in the Cosine beam [40] and HG beam [41] these changes in Cartesian coordinates with rectangular symmetry.

Ideal Cosine and Bessel beams formed as a result of plane waves and have constant intensity modulations in their transverse profile with fringe visibility value one. In a real case, it is not possible to generate a plane wave beam due to its infinite range of energy distribution, and depending upon the interfering laser beams' shape, we can generate different kinds of Cosine and Bessel beams. In real beams, the fringe visibility is spatially variant. The intensity distribution in Cosine beams can be modulated in a controlled manner through the interfering beams' polarization and spatial intensity distributions.

## 8. Conclusion

We successfully demonstrate the origin and characteristics of the Cosine beam with the aid of Dirac notation and ray optics. Our method of analysis was developed based on obliquely cross-propagating waves to produce a Cosine beam. The superposed waves are in standing waveform and propagating waveform in the respective directions of perpendicular and parallel to the optical axis of the Cosine beam. Our method of explanation can be easily applied to any kind of real Cosine beam by considering the necessary apodization function in the theoretical formulation. Further, we have shown that a series of optical needle array can be generated by the superposition of two 1D Cosine beams whose propagation direction is the same, but the planes of cosine modulations are orthogonal. The self-healing and non-diffraction nature of the optical needle array generated in Cosine beams can augment the quality of optical processes like particle trapping, bio-imaging, material processing, atomic guiding, etc.

The similarity between the Cosine beam, Bessel beam, and HG beams is provided with detailed analysis. The proposed method of analysis can have salient features in fundamental and applied optical processes.

## Supplemental data

### A. Two dimensional Cosine beam

The second kind Eq. of 2D Cosine beam is given by

$$|\Psi\rangle = |\psi_x\rangle|\psi_y\rangle$$
$$\Rightarrow |\Psi\rangle = e^{-ikz\cos\theta_x}\left(e^{-ikx\sin\theta_x}|\beta_x\rangle + e^{ikx\sin\theta_x}|\beta'_x\rangle\right) \times e^{-ikz\cos\theta_y}\left(e^{-iky\sin\theta_y}|\beta_y\rangle + e^{iky\sin\theta_y}|\beta'_y\rangle\right)$$
$$\Rightarrow |\Psi\rangle = e^{-ikz(\cos\theta_x+\cos\theta_y)}\left[e^{-ik(x\sin\theta_x+y\sin\theta_y)}|\beta_x\beta_y\rangle + e^{ik(x\sin\theta_x+y\sin\theta_y)}|\beta'_x\beta'_y\rangle\right.$$
$$\left. e^{-ik(x\sin\theta_x-y\sin\theta_y)}|\beta_x\beta'_y\rangle + e^{ik(x\sin\theta_x-y\sin\theta_y)}|\beta'_x\beta_y\rangle\right] \qquad (S_1)$$

The expressions for states $|\beta_i\rangle$ and $|\beta'_i\rangle$ are

$$|\beta_x\rangle = E(\cos\phi|a\rangle + \sin\phi\cos\theta_x|b\rangle + \sin\phi\sin\theta_x|c\rangle), \qquad (S_2 a)$$
$$|\beta'_x\rangle = E'(\cos\phi|a\rangle + \sin\phi\cos\theta_x|b\rangle + \sin\phi\sin\theta_x|c\rangle), \qquad (S_2 b)$$
$$|\beta_y\rangle = E(\cos\phi|a\rangle + \sin\phi\cos\theta_y|b\rangle + \sin\phi\sin\theta_y|c\rangle), \qquad (S_2 c)$$
$$|\beta'_y\rangle = E'(\cos\phi|a\rangle + \sin\phi\cos\theta_y|b\rangle + \sin\phi\sin\theta_y|c\rangle), \qquad (S_2 d)$$

and their product state is given by

$$|\beta_x\beta_y\rangle = E^2\xi, \qquad (S_3 a)$$
$$|\beta_x\beta'_y\rangle = |\beta'\beta\rangle = EE'\xi, \qquad (S_3 a)$$
$$|\beta'_x\beta'_y\rangle = E'^2\xi. \qquad (S_3 c)$$

Here,

$$\xi = \cos^2\phi|aa\rangle + \sin^2\phi\cos\theta_x\cos\theta_y|bb\rangle + \sin^2\phi\sin\theta_x\sin\theta_y|cc\rangle$$
$$+ \cos\phi\sin\phi\cos\theta_y|ab\rangle + \cos\phi\sin\phi\sin\theta_y|ac\rangle$$
$$+ \cos\phi\sin\phi\cos\theta_x|ba\rangle + \sin^2\phi\sin\theta_y\cos\theta_x|bc\rangle$$
$$+ \cos\phi\sin\phi\sin\theta_x|ca\rangle + \sin^2\phi\sin\theta_x\cos\theta_y|cb\rangle$$

$$\Rightarrow \xi = \cos^2\phi|aa\rangle + \sin^2\phi\cos\theta_x\cos\theta_y|bb\rangle + \sin^2\phi\sin\theta_x\sin\theta_y|cc\rangle$$
$$+ \cos\phi\sin\phi(\cos\theta_x + \cos\theta_y)|ab\rangle + \cos\phi\sin\phi(\sin\theta_x + \sin\theta_y)|ac\rangle \quad \because |ab\rangle = |ba\rangle$$
$$+ \sin^2\phi\sin(\theta_x + \theta_y)|bc\rangle \quad\quad (S_4)$$

and

$$|\xi|^2 = \cos^4\phi + \sin^4\phi\cos^2\theta_x\cos^2\theta_y + \sin^4\phi\sin^2\theta_x\sin^2\theta_y$$
$$+ \cos^2\phi\sin^2\phi(\cos\theta_x + \cos\theta_y)^2 + \cos^2\phi\sin^2\phi(\sin\theta_x + \sin\theta_y)^2$$
$$+ \sin^4\phi\sin^2(\theta_x + \theta_y) \quad\quad \because \langle a_1 a_2 | a_3 a_4 \rangle = \delta_{1,3}\delta_{2,4}$$

$$\Rightarrow |\xi|^2 = \cos^4\phi + \sin^4\phi(\cos^2\theta_x\cos^2\theta_y + \sin^2\theta_x\sin^2\theta_y)$$
$$+ \cos^2\phi\sin^2\phi((\cos\theta_x + \cos\theta_y)^2 + (\sin\theta_x + \sin\theta_y)^2)$$
$$+ \sin^4\phi\sin^2(\theta_x + \theta_y). \quad\quad (S_5)$$

With consideration of $E=E'$ we can obtain the following inner products

$$\langle \beta_x\beta_y | \beta_x\beta_y \rangle = \langle \beta_x\beta_y | \beta'_x\beta'_y \rangle = \langle \beta_x\beta_y | \beta'_x\beta_y \rangle = \langle \beta_x\beta_y | \beta_x\beta'_y \rangle = I^2|\xi|^2, \quad (S_6a)$$

$$\langle \beta'_x\beta_y | \beta_x\beta_y \rangle = \langle \beta'_x\beta_y | \beta'_x\beta'_y \rangle = \langle \beta'_x\beta_y | \beta'_x\beta_y \rangle = \langle \beta'_x\beta_y | \beta_x\beta'_y \rangle = I^2|\xi|^2, \quad (S_6b)$$

$$\langle \beta_x\beta'_y | \beta_x\beta_y \rangle = \langle \beta_x\beta'_y | \beta'_x\beta'_y \rangle = \langle \beta_x\beta'_y | \beta'_x\beta_y \rangle = \langle \beta_x\beta'_y | \beta_x\beta'_y \rangle = I^2|\xi|^2, \quad (S_6c)$$

$$\langle \beta'_x\beta'_y | \beta_x\beta_y \rangle = \langle \beta'_x\beta'_y | \beta'_x\beta'_y \rangle = \langle \beta'_x\beta'_y | \beta'_x\beta_y \rangle = \langle \beta'_x\beta'_y | \beta_x\beta'_y \rangle = I^2|\xi|^2. \quad (S_6d)$$

Finally, the 2D Cosine beam inner product is

$$\langle \Psi | \Psi \rangle = [e^{ik(x\sin\theta_x + y\sin\theta_y)}\langle \beta_x\beta_y| + e^{-ik(x\sin\theta_x + y\sin\theta_y)}\langle \beta'_x\beta'_y|$$
$$+ e^{-ik(x\sin\theta_x - y\sin\theta_y)}\langle \beta'_x\beta_y| + e^{ik(x\sin\theta_x - y\sin\theta_y)}\langle \beta_x\beta'_y|]$$
$$\times [e^{-ik(x\sin\theta_x + y\sin\theta_y)}|\beta_x\beta_y\rangle + e^{ik(x\sin\theta_x + y\sin\theta_y)}|\beta'_x\beta'_y\rangle$$
$$+ e^{ik(x\sin\theta_x - y\sin\theta_y)}|\beta'_x\beta_y\rangle + e^{-ik(x\sin\theta_x - y\sin\theta_y)}|\beta_x\beta'_y\rangle]$$

$$\Rightarrow \langle \Psi | \Psi \rangle = (\langle \beta_x\beta_y|\beta_x\beta_y\rangle + \langle \beta'_x\beta_y|\beta'_x\beta_y\rangle + \langle \beta_x\beta'_y|\beta_x\beta'_y\rangle + \langle \beta'_x\beta'_y|\beta'_x\beta'_y\rangle)$$
$$+ e^{2ik(x\sin\theta_x + y\sin\theta_y)}\langle \beta_x\beta_y|\beta'_x\beta'_y\rangle + e^{2ikx\sin\theta_x}\langle \beta_x\beta_y|\beta'_x\beta_y\rangle + e^{2iky\sin\theta_y}\langle \beta_x\beta_y|\beta_x\beta'_y\rangle$$
$$+ e^{-2ik(x\sin\theta_x + y\sin\theta_y)}\langle \beta'_x\beta'_y|\beta_x\beta_y\rangle + e^{-2iky\sin\theta_y}\langle \beta'_x\beta'_y|\beta'_x\beta_y\rangle + e^{-2ikx\sin\theta_x}\langle \beta'_x\beta'_y|\beta_x\beta'_y\rangle$$
$$+ e^{-2ikx\sin\theta_x}\langle \beta'_x\beta_y|\beta_x\beta_y\rangle + e^{2iky\sin\theta_y}\langle \beta'_x\beta_y|\beta'_x\beta'_y\rangle + e^{-2ik(x\sin\theta_x - y\sin\theta_y)}\langle \beta'_x\beta_y|\beta'_x\beta_y\rangle$$
$$+ e^{-2iky\sin\theta_y}\langle \beta_x\beta'_y|\beta_x\beta_y\rangle + e^{2ikx\sin\theta_x}\langle \beta_x\beta'_y|\beta'_x\beta'_y\rangle + e^{2ik(x\sin\theta_x - y\sin\theta_y)}\langle \beta_x\beta'_y|\beta'_x\beta_y\rangle$$

$$\Rightarrow \langle \Psi | \Psi \rangle = I^2|\xi|^2 (4 + e^{2ik(x\sin\theta_x + y\sin\theta_y)} + e^{2ikx\sin\theta_x} + e^{2iky\sin\theta_y}$$
$$+ e^{-2ik(x\sin\theta_x + y\sin\theta_y)} + e^{-2iky\sin\theta_y} + e^{-2ikx\sin\theta_x} + e^{-2ikx\sin\theta_x} + e^{2iky\sin\theta_y}$$
$$+ e^{-2ik(x\sin\theta_x - y\sin\theta_y)} + e^{-2iky\sin\theta_y} + e^{2ikx\sin\theta_x} + e^{2ik(x\sin\theta_x - y\sin\theta_y)})$$

$$\Rightarrow \langle\Psi|\Psi\rangle = I^2|\xi|^2\{4 + 2\cos[2k(x\sin\theta_x + y\sin\theta_y)] + 2\cos[2k(x\sin\theta_x - y\sin\theta_y)]$$
$$+ 4\cos(2kx\sin\theta_x) + 4\cos(2ky\sin\theta_y)\}$$
$$\Rightarrow \langle\Psi|\Psi\rangle = I^2|\xi|^2[4 + 4\cos(2kx\sin\theta_x)\cos(2ky\sin\theta_y) + 4\cos(2kx\sin\theta_x) + 4\cos(2ky\sin\theta_y)]$$
$$\Rightarrow \langle\Psi|\Psi\rangle = 4I^2|\xi|^2[1 + \cos(2kx\sin\theta_x)\cos(2ky\sin\theta_y) + \cos(2kx\sin\theta_x) + \cos(2ky\sin\theta_y))] \quad (S_7)$$

**B. Interference of two one dimensional Cosine beams with their plane of incidences are orthogonal to each other**

The intensity distribution of superposition of two 1D cosine beams is given by

$$\langle\psi|\psi\rangle = \left[e^{ikz\cos\theta_x}\left(e^{ikx\sin\theta_x}\langle\beta_x| + e^{-ikx\sin\theta_x}\langle\beta'_x|\right) + e^{ikz\cos\theta_y}\left(e^{iky\sin\theta_y}\langle\beta_y| + e^{-iky\sin\theta_y}\langle\beta'_y|\right)\right]$$
$$\left[e^{-ikz\cos\theta_x}\left(e^{-ikx\sin\theta_x}|\beta_x\rangle + e^{ikx\sin\theta_x}|\beta'_x\rangle\right) + e^{-ikz\cos\theta_y}\left(e^{-iky\sin\theta_y}|\beta_y\rangle + e^{iky\sin\theta_y}|\beta'_y\rangle\right)\right]$$

$$\Rightarrow \langle\psi|\psi\rangle = \left(\langle\beta_x|\beta_x\rangle + \langle\beta'_x|\beta'_x\rangle + e^{2ikx\sin\theta_x}\langle\beta_x|\beta'_x\rangle + e^{-2ikx\sin\theta_x}\langle\beta'_x|\beta_x\rangle\right)$$
$$+ \left(\langle\beta_y|\beta_y\rangle + \langle\beta'_y|\beta'_y\rangle + e^{2iky\sin\theta_y}\langle\beta_y|\beta'_y\rangle + e^{-2iky\sin\theta_y}\langle\beta'_y|\beta_y\rangle\right)$$
$$+ e^{ikz(\cos\theta_x - \cos\theta_y)}\begin{pmatrix} e^{ik(x\sin\theta_x - y\sin\theta_y)}\langle\beta_x|\beta_y\rangle + e^{ik(x\sin\theta_x + y\sin\theta_y)}\langle\beta_x|\beta'_y\rangle \\ + e^{-ik(x\sin\theta_x + y\sin\theta_y)}\langle\beta'_x|\beta_y\rangle + e^{-ik(x\sin\theta_x - y\sin\theta_y)}\langle\beta'_x|\beta'_y\rangle \end{pmatrix}$$
$$+ e^{-ikz(\cos\theta_x - \cos\theta_y)}\begin{pmatrix} e^{-ik(x\sin\theta_x - y\sin\theta_y)}\langle\beta_y|\beta_x\rangle + e^{ik(x\sin\theta_x + y\sin\theta_y)}\langle\beta_y|\beta'_x\rangle \\ + e^{-ik(x\sin\theta_x + y\sin\theta_y)}\langle\beta'_y|\beta_x\rangle + e^{ik(x\sin\theta_x - y\sin\theta_y)}\langle\beta'_y|\beta'_x\rangle \end{pmatrix}. \quad (S_8)$$

Here, inner product of states $|\beta\rangle$ and $|\beta'\rangle$ given by

$$\langle\beta_x|\beta_x\rangle = \langle\beta_y|\beta_y\rangle = |E|^2 \qquad (S_9a)$$
$$\langle\beta_x|\beta_y\rangle = \langle\beta_y|\beta_x\rangle = f(\phi,\theta_x,\theta_y)|E|^2 \qquad (S_9b)$$
$$\langle\beta'_x|\beta'_x\rangle = \langle\beta'_y|\beta'_y\rangle = |E'|^2 \qquad (S_9c)$$
$$\langle\beta'_x|\beta'_y\rangle = \langle\beta'_y|\beta'_x\rangle = f(\phi,\theta_x,\theta_y)|E'|^2 \qquad (S_9d)$$
$$\langle\beta_x|\beta'_x\rangle = \langle\beta_y|\beta'_y\rangle = E^*E' \qquad (S_9e)$$
$$\langle\beta_x|\beta'_y\rangle = \langle\beta_y|\beta'_x\rangle = f(\phi,\theta_x,\theta_y)E^*E' \qquad (S_9f)$$
$$\langle\beta'_x|\beta_x\rangle = \langle\beta'_y|\beta_y\rangle = EE'^* \qquad (S_9g)$$
$$\langle\beta'_y|\beta_x\rangle = \langle\beta'_x|\beta_y\rangle = f(\phi,\theta_x,\theta_y)EE'^* \qquad (S_9h)$$

with

$$f(\phi,\theta_x,\theta_y) = \cos^2\phi\cos\theta_x + \sin^2\phi\cos\theta_y + \cos\phi\sin\phi\sin\theta_x\sin\theta_y.$$

The Eq. $S_8$ can be further simplified with Eqs. $S_9(a-h)$ as

$$\langle\psi|\psi\rangle = \left(|E|^2 + |E'|^2 + E^*E'e^{2ikx\sin\theta_x} + EE'^*e^{-2ikx\sin\theta_x}\right) + \left(|E|^2 + |E'|^2 + E^*E'e^{2iky\sin\theta_y} + EE'^*e^{-2iky\sin\theta_y}\right)$$
$$+ \left[e^{ikz(\cos\theta_x - \cos\theta_y)}\left(e^{ik(x\sin\theta_x - y\sin\theta_y)}|E|^2 + e^{ik(x\sin\theta_x + y\sin\theta_y)}E^*E' + e^{-ik(x\sin\theta_x + y\sin\theta_y)}EE'^* + e^{-ik(x\sin\theta_x - y\sin\theta_y)}|E'|^2\right)\right.$$
$$\left. + e^{-ikz(\cos\theta_x - \cos\theta_y)}\left(e^{-ik(x\sin\theta_x - y\sin\theta_y)}|E|^2 + e^{ik(x\sin\theta_x + y\sin\theta_y)}E^*E' + e^{-ik(x\sin\theta_x + y\sin\theta_y)}EE'^* + e^{ik(x\sin\theta_x - y\sin\theta_y)}|E'|^2\right)\right]$$
$$\left(\cos^2\phi\cos\theta_x + \sin^2\phi\cos\theta_y + \cos\phi\sin\phi\sin\theta_x\sin\theta_y\right)$$

$$\Rightarrow \langle\psi|\psi\rangle = \left[I + I' + \sqrt{II'}\left(e^{2ikx\sin\theta_x} + e^{-2ikx\sin\theta_x}\right)\right] + \left[I + I' + \sqrt{II'}\left(e^{2iky\sin\theta_y} + e^{-2iky\sin\theta_y}\right)\right]$$
$$+ \left[e^{ikz(\cos\theta_x - \cos\theta_y)}\left(e^{ik(x\sin\theta_x - y\sin\theta_y)}I + e^{ik(x\sin\theta_x + y\sin\theta_y)}\sqrt{II'} + e^{-ik(x\sin\theta_x + y\sin\theta_y)}\sqrt{II'} + e^{-ik(x\sin\theta_x - y\sin\theta_y)}I'\right)\right.$$
$$\left. + e^{-ikz(\cos\theta_x - \cos\theta_y)}\left(e^{-ik(x\sin\theta_x - y\sin\theta_y)}I + e^{ik(x\sin\theta_x + y\sin\theta_y)}\sqrt{II'} + e^{-ik(x\sin\theta_x + y\sin\theta_y)}\sqrt{II'} + e^{ik(x\sin\theta_x - y\sin\theta_y)}I'\right)\right]$$
$$\left(\cos^2\phi\cos\theta_x + \sin^2\phi\cos\theta_y + \cos\phi\sin\phi\sin\theta_x\sin\theta_y\right)$$

$$\Rightarrow \langle\psi|\psi\rangle = [2I + 2I\cos(2kx\sin\theta_x)] + [2I + 2I\cos(2ky\sin\theta_y)]$$
$$+ I\left[e^{ikz(\cos\theta_x - \cos\theta_y)}\left(e^{ik(x\sin\theta_x - y\sin\theta_y)} + e^{ik(x\sin\theta_x + y\sin\theta_y)} + e^{-ik(x\sin\theta_x + y\sin\theta_y)} + e^{-ik(x\sin\theta_x - y\sin\theta_y)}\right)\right.$$
$$\left. + e^{-ikz(\cos\theta_x - \cos\theta_y)}\left(e^{-ik(x\sin\theta_x - y\sin\theta_y)} + e^{ik(x\sin\theta_x + y\sin\theta_y)} + e^{-ik(x\sin\theta_x + y\sin\theta_y)} + e^{ik(x\sin\theta_x - y\sin\theta_y)}\right)\right]$$
$$(\cos^2\phi\cos\theta_x + \sin^2\phi\cos\theta_y + \cos\phi\sin\phi\sin\theta_x\sin\theta_y)$$

$$\Rightarrow \langle\psi|\psi\rangle = 4I[\cos^2(kx\sin\theta_x) + \cos^2(ky\sin\theta_y)]$$
$$+ I\left[e^{ikz(\cos\theta_x - \cos\theta_y)}(2\cos(kx\sin\theta_x - ky\sin\theta_y) + 2\cos(kx\sin\theta_x + ky\sin\theta_y))\right.$$
$$\left. + e^{-ikz(\cos\theta_x - \cos\theta_y)}(2\cos(kx\sin\theta_x - ky\sin\theta_y) + 2\cos(kx\sin\theta_x + ky\sin\theta_y))\right]$$
$$(\cos^2\phi\cos\theta_x + \sin^2\phi\cos\theta_y + \cos\phi\sin\phi\sin\theta_x\sin\theta_y)$$

$$\Rightarrow \langle\psi|\psi\rangle = 4I[\cos^2(kx\sin\theta_x) + \cos^2(ky\sin\theta_y)]$$
$$+ 2I\left[e^{ikz(\cos\theta_x - \cos\theta_y)} + e^{-ikz(\cos\theta_x - \cos\theta_y)}\right](\cos(kx\sin\theta_x - ky\sin\theta_y) + \cos(kx\sin\theta_x + ky\sin\theta_y))$$
$$(\cos^2\phi\cos\theta_x + \sin^2\phi\cos\theta_y + \cos\phi\sin\phi\sin\theta_x\sin\theta_y)$$

$$\Rightarrow \langle\psi|\psi\rangle = 4I[\cos^2(kx\sin\theta_x) + \cos^2(ky\sin\theta_y)]$$
$$+ 4I\cos(kz\cos\theta_x - kz\cos\theta_y)(\cos(kx\sin\theta_x - ky\sin\theta_y) + \cos(kx\sin\theta_x + ky\sin\theta_y))$$
$$(\cos^2\phi\cos\theta_x + \sin^2\phi\cos\theta_y + \cos\phi\sin\phi\sin\theta_x\sin\theta_y)$$

$$\langle\psi|\psi\rangle = 4I[\cos^2(kx\sin\theta_x) + \cos^2(ky\sin\theta_y)] + 8I\cos(kz\cos\theta_x - kz\cos\theta_y)\cos(kx\sin\theta_x)\cos(ky\sin\theta_y)$$
$$(\cos^2\phi\cos\theta_x + \sin^2\phi\cos\theta_y + \cos\phi\sin\phi\sin\theta_x\sin\theta_y). \qquad (S_{10})$$

**C. Superposition of orthogonally polarized two one dimensional Cosine beams**

Eq. $S_{10}$ can be further simplified for the special cases given below

*$C_1$. Linearly Polarized one dimensional Cosine beams with their polarization in their own plane of incident*

The state of superposition can be written in terms of interfering beams as

$$|\psi\rangle = e^{-ikz\cos\theta_x}\left(e^{-ikx\sin\theta_x}|\beta_x\rangle + e^{ikx\sin\theta_x}|\beta'_x\rangle\right) + e^{-ikz\cos\theta_y}\left(e^{-iky\sin\theta_y}|\beta_y\rangle + e^{iky\sin\theta_y}|\beta'_y\rangle\right) \qquad (S_{11})$$

with its three dimensional polarization states $|\beta\rangle$ and $|\beta'\rangle$ have simpler form as

$$|\beta_x\rangle = E(\cos\theta_x|a\rangle + \sin\theta_x|c\rangle), \qquad (S_{12}a)$$
$$|\beta'_x\rangle = E'(\cos\theta_x|a\rangle + \sin\theta_x|c\rangle), \qquad (S_{12}b)$$
$$|\beta_y\rangle = E(\cos\theta_y|b\rangle + \sin\theta_y|c\rangle), \qquad (S_{12}c)$$
$$|\beta'_y\rangle = E'(\cos\theta_y|b\rangle + \sin\theta_y|c\rangle), \qquad (S_{12}d)$$

and their inner product given by

$$\langle\beta_x|\beta_x\rangle = \langle\beta_y|\beta_y\rangle = |E|^2 \qquad (S_{13}a)$$
$$\langle\beta_x|\beta_y\rangle = \langle\beta_y|\beta_x\rangle = \sin\theta_x\sin\theta_y|E|^2 \qquad (S_{13}b)$$
$$\langle\beta'_x|\beta'_x\rangle = \langle\beta'_y|\beta'_y\rangle = |E'|^2 \qquad (S_{13}c)$$
$$\langle\beta'_x|\beta'_y\rangle = \langle\beta'_y|\beta'_x\rangle = \sin\theta_x\sin\theta_y|E'|^2 \qquad (S_{13}d)$$
$$\langle\beta_x|\beta'_x\rangle = \langle\beta_y|\beta'_y\rangle = E^*E' \qquad (S_{13}e)$$
$$\langle\beta_x|\beta'_y\rangle = \langle\beta_y|\beta'_x\rangle = \sin\theta_x\sin\theta_y E^*E' \qquad (S_{13}f)$$
$$\langle\beta'_x|\beta_x\rangle = \langle\beta'_y|\beta_y\rangle = EE^* \qquad (S_{13}g)$$
$$\langle\beta'_y|\beta_x\rangle = \langle\beta'_x|\beta_y\rangle = \sin\theta_x\sin\theta_y EE^* \qquad (S_{13}h)$$

The intensity distribution in the superposition is given by

$$\langle\psi|\psi\rangle = \left[e^{ikz\cos\theta_x}\left(e^{ikx\sin\theta_x}\langle\beta_x| + e^{-ikx\sin\theta_x}\langle\beta'_x|\right) + e^{ikz\cos\theta_y}\left(e^{iky\sin\theta_y}\langle\beta_y| + e^{-iky\sin\theta_y}\langle\beta'_y|\right)\right]$$
$$\left[e^{-ikz\cos\theta_x}\left(e^{-ikx\sin\theta_x}|\beta_x\rangle + e^{ikx\sin\theta_x}|\beta'_x\rangle\right) + e^{-ikz\cos\theta_y}\left(e^{-iky\sin\theta_y}|\beta_y\rangle + e^{iky\sin\theta_y}|\beta'_y\rangle\right)\right]$$

$$\Rightarrow \langle\psi|\psi\rangle = \left(\langle\beta_x|\beta_x\rangle + \langle\beta'_x|\beta'_x\rangle + e^{2ikx\sin\theta_x}\langle\beta_x|\beta'_x\rangle + e^{-2ikx\sin\theta_x}\langle\beta'_x|\beta_x\rangle\right)$$
$$+ \left(\langle\beta_y|\beta_y\rangle + \langle\beta'_y|\beta'_y\rangle + e^{2iky\sin\theta_y}\langle\beta_y|\beta'_y\rangle + e^{-2iky\sin\theta_y}\langle\beta'_y|\beta_y\rangle\right)$$

with consideration of $I=I'$

$$\langle\psi|\psi\rangle = 4I[\cos^2(kx\sin\theta_x) + \cos^2(ky\sin\theta_y)]$$
$$+ 8I\cos(kz\cos\theta_x - kz\cos\theta_y)\cos(kx\sin\theta_x)\cos(ky\sin\theta_y)\sin\theta_x\sin\theta_y, \quad (S_{14})$$

$$\langle\psi|\psi\rangle = 4I[\cos^2(kx\sin\theta) + \cos^2(ky\sin\theta)]$$
$$+ 8I\cos(kx\sin\theta)\cos(ky\sin\theta)\sin^2\theta \quad (\forall\,\theta_x = \theta_y). \quad (S_{15})$$

**C₂. Linearly Polarized one dimensional Cosine beams with their polarization is perpendicular to their own plane of incident**

The state of superposition beams is same as $S_{11}$ and its polarization states are

$|\beta_x\rangle = |b\rangle E,$                 $(S_{16}a)$

$|\beta'_x\rangle = |b\rangle E',$                $(S_{16}b)$

$|\beta_y\rangle = |a\rangle E,$                 $(S_{16}c)$

$|\beta'_y\rangle = |a\rangle E',$                $(S_{16}d)$

and their inner products are

$\langle\beta_x|\beta_x\rangle = \langle\beta_y|\beta_y\rangle = |E|^2,$          $(S_{17}a)$

$\langle\beta_x|\beta_y\rangle = \langle\beta_y|\beta_x\rangle = 0,$          $(S_{17}b)$

$\langle\beta'_x|\beta'_x\rangle = \langle\beta'_y|\beta'_y\rangle = |E'|^2,$          $(S_{17}c)$

$\langle\beta'_x|\beta'_y\rangle = \langle\beta'_y|\beta'_x\rangle = 0,$          $(S_{17}d)$

$\langle\beta_x|\beta'_x\rangle = \langle\beta_y|\beta'_y\rangle = E^*E',$          $(S_{17}e)$

$\langle\beta_x|\beta'_y\rangle = \langle\beta_y|\beta'_x\rangle = 0,$          $(S_{17}f)$

$\langle\beta'_x|\beta_x\rangle = \langle\beta'_y|\beta_y\rangle = EE'^*,$          $(S_{17}g)$

$\langle\beta'_y|\beta_x\rangle = \langle\beta'_x|\beta_y\rangle = 0.$          $(S_{17}h)$

The intensity distribution in the interference is given by

$$\langle\psi|\psi\rangle = \left[e^{ikz\cos\theta_x}\left(e^{ikx\sin\theta_x}\langle\beta_x| + e^{-ikx\sin\theta_x}\langle\beta'_x|\right) + e^{ikz\cos\theta_y}\left(e^{iky\sin\theta_y}\langle\beta_y| + e^{-iky\sin\theta_y}\langle\beta'_y|\right)\right]$$
$$\left[e^{-ikz\cos\theta_x}\left(e^{-ikx\sin\theta_x}|\beta_x\rangle + e^{ikx\sin\theta_x}|\beta'_x\rangle\right) + e^{-ikz\cos\theta_y}\left(e^{-iky\sin\theta_y}|\beta_y\rangle + e^{iky\sin\theta_y}|\beta'_y\rangle\right)\right]$$

$$\Rightarrow \langle\psi|\psi\rangle = \left(\langle\beta_x|\beta_x\rangle + \langle\beta'_x|\beta'_x\rangle + e^{2ikx\sin\theta_x}\langle\beta_x|\beta'_x\rangle + e^{-2ikx\sin\theta_x}\langle\beta'_x|\beta_x\rangle\right)$$
$$+ \left(\langle\beta_y|\beta_y\rangle + \langle\beta'_y|\beta'_y\rangle + e^{2iky\sin\theta_y}\langle\beta_y|\beta'_y\rangle + e^{-2iky\sin\theta_y}\langle\beta'_y|\beta_y\rangle\right)$$

$$\Rightarrow \langle\psi|\psi\rangle = \left(|E|^2 + |E'|^2 + E^*E'e^{2ikx\sin\theta_x} + EE'^*e^{-2ikx\sin\theta_x}\right) + \left(|E|^2 + |E'|^2 + E^*E'e^{2iky\sin\theta_y} + EE'^*e^{-2iky\sin\theta_y}\right)$$

$$\Rightarrow \langle\psi|\psi\rangle = \left[I + I' + \sqrt{II'}\left(e^{2ikx\sin\theta_x} + e^{-2ikx\sin\theta_x}\right)\right] + \left[I + I' + \sqrt{II'}\left(e^{2iky\sin\theta_y} + e^{-2iky\sin\theta_y}\right)\right]$$

with consideration of $I=I'$

$$\Rightarrow \langle\psi|\psi\rangle = [2I + 2I\cos(2kx\sin\theta_x)] + [2I + 2I\cos(2ky\sin\theta_y)]$$

$$\Rightarrow \langle\psi|\psi\rangle = 4I[\cos^2(kx\sin\theta_x) + \cos^2(ky\sin\theta_y)] \qquad (S_{18})$$